\documentclass{raa}            

\usepackage{graphicx,times}             

\begin{document}

 \title{On the possible wind nebula of magnetar Swift J1834.9$-$0846: a magnetism-powered synchrotron nebula
}

   \volnopage{Vol.0 (200x) No.0, 000--000}      
   \setcounter{page}{1}          

     \author{H. Tong
      \inst{}
      }

   \institute{Xinjiang Astronomical Observatory, Chinese Academy of Sciences, Urumqi, Xinjiang 830011,
    China; {\it tonghao@xao.ac.cn}\\
      }

   \date{Received~~2012 month day; accepted~~2012~~month day}

\abstract{ 
Recently, the magnetar Swift J1834.9$-$0846 is reported to have a possible wind nebula. It is shown that both the 
magnetar and its wind nebula are understandable in the wind braking scenario. The magnetar's rotational energy loss 
rate is not enough to power the particle luminosity. The required particle luminosity should be about $10^{36} \,\rm erg \,s^{-1}$ to $10^{38} \,\rm erg \,s^{-1}$. 
It is obtained in three different approaches: considering wind braking of Swift J1834.9$-$0846; the spectral and spatial observations of the wind nebula; and an empirical upper bound on wind nebula X-ray luminosity. The nebula magnetic field 
is about $10^{-4} \,\rm G$. The possible wind nebula of Swift J1834.9$-$0846 should be a magnetar wind nebula. 
It is powered by the magnetic energy release of the magnetar. 
\keywords{pulsars: individual (Swift J1834.9$-$0846)---stars:magnetars---stars:neutron\\ winds and outflows}
}

   \authorrunning{H. Tong}            
   \titlerunning{On the possible wind nebula of magnetar Swift J1834.9$-$0846}  

   \maketitle

%
%
\section{Introduction}

Magnetars are a special kind of isolated neutron stars. They may be neutron stars powered by their strong 
magnetic field (Duncan \& Thompson 1992; Mereghetti et al. 2015). The magnetic energy release will 
dominate the observational behaviors of the magnetar's persistent X-ray luminosity, bursts, and spin-down. 
During the outbursts, the magnetar X-ray luminosity will decrease, accompanied by a decaying spin-down rate and/or
other timing events (Scholz et al. 2014; Younes et al. 2015; Pintore et al. 2016; Tong \& Xu 2013; Tong 2014). 
The variable spin-down rate
of magnetars may be due a strong particle wind flowing out from the magnetar magnetosphere (Tong et al. 2013).
This scenario is called ``wind braking of magnetars''. In the case of normal pulsars, the particle outflow is powered
by the rotational energy of the central neutron stars. This particle outflow will finally be seen as a wind nebula around 
the pulsar (Gaensler \& Slane 2006; Kargaltsev et al. 2015). Analogous with normal pulsars, wind braking of
magnetars predicted the existence of a magnetar wind nebula (Tong et al. 2013). The difference between a magnetar 
wind nebula (MWN) and a traditional pulsar wind nebula (PWN) is that the magnetar wind nebula can be powered by the magnetic energy release. The central neutron star's magnetic energy may first be converted to a system of non-thermal particles. These particle will affect the radiation spectra of the magnetar, contribute a braking torque to the magnetar. 
When they flow out to the surroundings, they may be seen in the form of wind nebula. 

However, the observations of magnetar wind nebula were not conclusive. For example, the extended emission around 
magnetar Swift J1834.9$-$0846 was thought to be a dust scattering halo (Kargaltsev et al. 2012). Later, Younes et al. 
(2012) proposed that some component may be due to a wind nebula. However, Esposito et al. (2013) used more 
observations to say that it should be a dust scattering halo. 
Recently, Younes et al. (2016) confirmed the idea of a wind nebula using two new deep XMM-Newton observations.  
With the addition of recent observations, the extended emission around Swift J1834.9$-$0846 are likely to be 
a wind nebula. In the following, we will calculate the properties of both the magnetar and its possible wind nebula. 
This may be the foundation of future more detailed studies. 

\section{Wind braking of magnetar Swift J1834.9$-$0846}

The magnetar Swift J1834.9$-$0846 has a rotational period $P=2.48\,\rm s$ and period derivative 
$\dot{P}=7.96 \times 10^{-12} \,\rm s\,s^{-1}$ (Kargaltsev et al. 2012). 
For a typical neutron star moment of inertia of $10^{45}\,\rm g \,cm^2$, 
the corresponding rotational energy loss rate is $\dot{E}_{\rm rot} =2.1 \times 10^{34} \,\rm erg \, s^{-1}$. 
Assuming magnetic dipole braking in vacuum, the magnetar characteristic age is 
$\tau_{\rm c} = \frac{P}{2\dot{P}} =4.9 \,\rm kyr$, the characteristic magnetic field at the magnetic pole is 
$B_{\rm c} = 6.4\times 10^{19} \sqrt{P \dot{P}} =2.8\times 10^{14} \,\rm G$. Swift J1834.9$-$0846 may be associated with the supernova remnant W41 (Tian et al. 2007; Kargaltsev et al. 2012). The distance and age provided by the supernova remnant 
is about $4\,\rm kpc$ and $60$-$200\,\rm kyr$, respectively (Tian et al. 2007). Assuming a distance of $4\,\rm kpc$, the peak X-ray luminosity can be as high as $10^{35}\, \rm erg \,s^{-1}$ during outburst, which quickly decreases with time (Kargaltsev et al. 2012). 
The quiescent X-ray luminosity of Swift J1834.9$-$0846 can be more than $10^3$ times fainter (Kargaltsev et al. 2012; Younes et al. 2016). Considering that only a small portion of the particle energy is converted to non-thermal X-rays, the particle luminosity 
of Swift J1834.9$-$0846 can be as high as $10^{36} \,\rm erg \,s^{-1}$ or higher, which depends on the X-ray efficiency. 
This particle luminosity is much higher than the rotational energy loss rate. Then, the spin-down of Swift J1834.9$-$0846 will be dominated by the particle wind (Tong et al. 2013 and references therein). 
Applying the wind braking model to Swift J1834.9$-$0846, its dipole magnetic field and age can be obtained. 

In the case of wind braking of magnetars, the corresponding magnetic field at the magnetic pole is (Tong et al. 2013, equation (31))
\begin{equation}
B_0 = 4.0\times 10^{25} \frac{\dot{P}}{P} L_{\rm p,35}^{-1/2} \,\rm G,
\end{equation} 
where $L_{\rm p,35}$ is the particle luminosity in units of $10^{35} \,\rm erg\,s^{-1}$. For a particle luminosity of $10^{36} \,\rm erg\,s^{-1}$, the polar magnetic field of Swift J1834.9$-$0846 is $B_0 =4.1\times 10^{13} \,\rm G$. The rotational energy loss rate 
due to magnetic dipole braking is proportional to $\Omega^4$, where $\Omega$ is the angular velocity of the neutron star. 
While, the rotational energy loss rate due to wind braking is proportional to $\Omega^2$. Therefore, the magnetar will mainly be spun down by magnetic dipole radiation during its early age. Later, when it has slowed down significantly, it will enter into the wind braking domain. 
For Swift J1834.9$-$0846, the transition period between magnetic dipole braking and wind braking is $P_1 =0.36 \,\rm s$ (Tong et al. 2013, equation (36)). The age of Swift J1834.9$-$0846 in the wind braking model is (Tong et al. 2013, equation(41))
\begin{equation}
t = (1+ 2 \log \frac{P}{P_1}) \, \tau_{\rm c}.
\end{equation}
The age of Swift J1834.9$-$0846 is about $20 \,\rm kyr$. 
For a higher particle luminosity of $10^{38} \,\rm erg \, s^{-1}$, by repeating the above calculations, the magnetar age is about $50 \,\rm kyr$. The corresponding supernova remnant age is about $60$-$200\,\rm kyr$ (Tian et al. 2007; Castro et al. 2013). 
The uncertainties in the measured supernova remnant age may be a factor of several. The age of  Swift J1834.9$-$0846 in the wind braking model may be view as consistent with supernova remnant age. Therefore, by applying the wind braking model to Swift J1834.9$-$0846, the age  discrepancy between the magnetar characteristic age and the corresponding supernova remnant age can be solved. The requirement is that the particle luminosity of Swift J1834.9$-$0846 is about $10^{36} \,\rm erg \,s^{-1}$ to $10^{38} \,\rm erg \,s^{-1}$.  It is also possible that the particle luminosity has a finite duty cycle and is relatively high temporarily. This will result in an enhanced spin-down rate. The corresponding characteristic age will be much smaller than the true age. This will also contribute to solution of the age discrepancy. 

\section{Diagnosing the possible wind nebula}

The possible  wind nebula around Swift J1834.9$-$0846 is observed in the soft X-ray range, between $2$-$10 \,\rm keV$ (Younes et al. 2016). Its angular radius is about $100''$. For a distance of $4\,\rm kpc$, the radius of the wind nebula is about $2 \,\rm pc$. 
Considering the asymmetrical shape of the wind nebula, a spherical radius of $1 \,\rm pc$ is adopted in the following calculations. 
Assuming a synchrotron origin for the X-ray emission, the typical synchrotron photon energy is (Ghisellini 2013)
\begin{equation}\label{Ephoton}
h \gamma^2 \frac{e B}{2 \pi m_e c} \approx 5 \,\rm keV,
\end{equation}
where $h$ is Planck's constant, $e$ is the absolute electron charge, $m_e$ is the electron rest mass, $c$ is the speed of light, 
$\gamma$ is the Lorentz factor of electrons responsible for the soft X-ray emissions, and $B$ is the nebula magnetic field in the emission region. A typical photon energy of $5\,\rm keV$ is adopted. There are possible signature of spectra softening with radius in the wind nebula (Younes et al. 2016).  This may be due to synchrotron burnoff or absorption effect . If due to synchrotron burnoff, it will require that the synchrotron cooling time scale is smaller than the crossing time of the nebula region (Gaensler \& Slane 2006; Chen et al. 2006). For a flow velocity of $c/3$ (Kennel \& Coroniti 1984), the crossing time is about $10$ years for a nebula with radius of $1 \,\rm pc$. Then, the synchrotron cooling time scale is (Ghisellini 2013)
\begin{equation}\label{tsyn}
\frac{24.57}{B^2 \gamma}\,\rm yr \leq 10 \,\rm yr.
\end{equation}
 Solving equations (\ref{Ephoton}) and (\ref{tsyn}) simultaneously, the nebula magnetic field is $B \geq 2.4 \times 10^{-4} \,\rm G$
 and the electron Lorentz factor is $\gamma \leq 4.2\times 10^7$. 

The nebula photon index is about $\Gamma \approx 2$ (Younes et al. 2016). The underlying electron distribution may also have
a power law form $N(\gamma) =K \gamma^{-p}$.  $N(\gamma){\rm d} \gamma$ is the number of electrons per cubic centimeter in the energy range $\gamma$-$\gamma +{\rm d}\gamma$, $K$ is the normalization in units of $\rm cm^{-3}$, and $p$ is the power law index. The photon index of $\Gamma \approx 2$ means that the electron power law index is $p=2\Gamma-1 \approx 3$. 
A Fermi type shock acceleration will generally result in a particle power law index of $p\approx 2$ (Rosswog \& Bruggen 2007). 
The corresponding photon index of synchrotron emission is $\Gamma=\frac{p+1}{2} \approx 1.5$. 
This may explain the photon index in the inner region of the wind nebula: $\Gamma_{\rm Inn} =1.3\pm 0.3$ (Younes et al. 2016). 
Synchrotron cooling etc will result in a steeper particle spectrum. This may correspond to the steeper 
photon index in the outer region of the wind nebula. 
At a distance of $4\,\rm kpc$, the nebula X-ray luminosity is about $L(0.5-10\,\rm keV) \approx 2.5\times 10^{33} \,\rm erg \, s^{-1}$ (Younes et al. 2016). 
The synchrotron power of a single electron is $P_{\rm s} = \frac{4}{3} \sigma_{\rm T} c U_{\rm B} \gamma^2$, where $\sigma_{\rm T}$ is the Thomson cross section, $U_{\rm B}$ is the magnetic energy density in the emission region (Ghisellini 2013). The nebula X-ray luminosity can be approximated as 
\begin{equation}
L_{\rm nebula} = P_{\rm s} N(\gamma) \gamma V_{\rm nebula}, 
\end{equation}
where $V_{\rm nebula}$ is the volume of the nebula. From this equation, the normalization $K$ is found to be about $K \approx 0.33 \,\rm cm^{-3}$. The energy density of the high energy electrons is about $6.4\times 10^{-15} \,\rm erg \,cm^{-3}$. 

The magnetic energy density in the emission region is about $U_{\rm B} =\frac{B^2}{8\pi} \approx 2.3\times 10^{-9} \,\rm erg \,cm^{-3}$. Assuming pressure balance between the magnetic field and the ambient material, the ambient pressure of the nebula 
should also be about $P_{\rm amb} \sim 10^{-9} \,\rm erg \,cm^{-3}$. The particle wind becomes near isotropic far away from the magnetar (Tong et al. 2013). The termination shock radius of the wind nebula is (Gaensler \& Slane 2006)
\begin{equation}
R_{\rm w} =\sqrt{\frac{L_{\rm p}}{4\pi c P_{\rm amb}}} =0.02 \, L_{\rm p,36}^{1/2} P_{\rm amb, -9}^{-1/2} \, \rm pc, 
\end{equation}
where $L_{p,36}$ is the particle luminosity in units of $10^{36} \,\rm erg \,s^{-1}$ and $P_{\rm amb, -9}$ is the ambient pressure
in units of $10^{-9} \,\rm erg \,cm^{-3}$. For a particle luminosity of $10^{36} \,\rm erg \,s^{-1}$, the termination shock radius 
 is about $0.02 \,\rm pc$. If the particle luminosity is as high as $10^{38} \,\rm erg \,s^{-1}$, the termination shock radius 
is about $0.2 \,\rm pc$. The wind nebula radius can be ten times larger than the termination shock radius (Table 2 in Kargaltsev \& Pavlov 2008). Therefore, the wind nebula can be produced if the particle luminosity is about $10^{36} \,\rm erg \,s^{-1}$ to 
$10^{38} \,\rm erg \,s^{-1}$. This is consistent with the above spin-down analysis of the central magnetar. A wind luminosity of 
$10^{36} \,\rm erg \,s^{-1}$ or higher is much larger than the central magnetar's rotational energy loss rate (about $2.1\times 10^{34} \,\rm erg \,s^{-1}$). Therefore, the possible 
wind nebula around Swift J1834.9$-$0846 can not be rotation-powered. It should be a wind nebula powered by the magnetic energy release of the central magnetar, i.e. a magnetar wind nebula. For this possible magnetar wind nebula, its X-ray efficiency is $\eta =L_{\rm x}/L_{\rm p} \sim 10^{-5}$-$10^{-3}$. It is similar to the X-ray efficiency of  pulsar wind nebula (Kargaltsev \& Pavlov 2008). 

The supernova remnants associated with magnetars are similar to those of normal pulsars (Vink \& Kuiper 2006; Martin et al. 2014). For a typical supernova remnant, the reverse shock will reach the supernova remnant centre after about $7\,\rm kyr$ (Reynolds \& Chevalier 1984). At this time, the reverse shock will crush the previous wind nebula (Gaensler \& Slane 2006). The wind nebula radius will decrease and reexpand later (Gelfand et al. 2009). The wind nebula around Swift J1834.9$-$0846 may be older than $20 \,\rm kyr$ for particle luminosity higher than $10^{36} \,\rm erg \,s^{-1}$.  The corresponding supernova remnant may already entered into the radiative phase. Therefore, it may be a wind nebula expanding in a radiative phase supernova remnant.

\section{Empirical upper bound on nebula X-ray luminosity}

The wind nebula luminosity is constant over 9 years time (Younes et al. 2016). For a particle luminosity of $10^{36} \,\rm erg\, s^{-1}$ to $10^{38} \,\rm erg\, s^{-1}$, the total energy released is about $10^{44} \,\rm erg$ to $10^{46} \,\rm erg$. This is equivalent to the energy released during magnetar giant flares (Mereghetti 2008). 
However, no giant flare is observed from Swift J1834.9$-$0846. 
And the spin-down rate of Swift J1834.9$-$0846 is relatively stable (Esposito et al. 2013). Therefore, the particle wind from Swift J1834.9$-$0846 should be relatively stable in the past years. The particle luminosity may also contain a burst component (Harding et al. 1999; Tong et al. 2013). For a bursting particle luminosity of $10^{37} \,\rm erg \,s^{-1}$, with a duty cycle of $0.1$,  it may have the same effect of a persistent particle luminosity of $10^{36} \,\rm erg \,s^{-1}$ (Tong et al. 2013). An empirical upper bound on the wind nebula X-ray luminosity is: $\log L_{\rm x} < 1.6 \log \dot{E}-24.2$ (Kargaltsev \& Pavlov 2008). Here $\dot{E}$ is the particle luminosity. It is the rotational energy loss rate in the case of normal pulsars.  For a particle luminosity of $10^{36} \,\rm erg \,s^{-1}$, the empirical upper bound on the wind nebula X-ray luminosity is about $2.5\times 10^{33} \,\rm erg\, s^{-1}$. The observed wind nebula X-ray luminosity around Swift J1834.9$-$0846 is also about $2.5 \times 10^{33} \,\rm erg \,s^{-1}$. Therefore, empirically, the particle wind luminosity of Swift J1834.9$-$0846 should be $10^{36} \,\rm erg \,s^{-1}$ or higher.  This is consistent with the above analysis. 

\section{Discussion and conclusion}

There are possible non-detection of a wind nebula in magnetar magnetar 1E 1547.0$-$5408 (Olausen et al. 2011) and  magnetar SGR 1806$-$20 (Vigano et al. 2014). Possible signatures of wind nebula are reported in the high magnetic field 
rotating radio transients RRAT J1819$-$1458 (Camero-Arranz et al. 2013), and magnear SGR J1935$+$2154 (Isreal et al. 2016). 
Both the non-detection and possible detection needs further studies. After some timing events of magnetars , upper limits on possible extended emission are also reported (Archibald et al. 2013; Scholz et al. 2014). Typically, the supposed extended emission have 
less than $2\%$ the magnetar's X-ray luminosity. However, the X-ray efficiency of known wind nebula is about $10^{-4}$. Therefore, 
a $2\%$ upper limit is not constraining (Tong 2015). 

The possible wind nebula around Swift J1834.9$-$0846 also needs further confirmations. Its association with supernova remnant W41 
is questionable (Tian et al. 2007; Kargaltsev et al. 2012; Castro et al. 2013). For the current observations (Younes et al. 2016), a wind luminosity of 
$10^{36} \,\rm erg\,s^{-1}$ to $10^{38} \,\rm erg \,s^{-1}$ and a nebula magnetic field of $\sim 10^{-4} \,\rm G$ is required. 
The particle wind may be due to persistent magnetic activities induced by the strong internal magnetic field (Thompson \& Duancan 1995; Tong et al. 2013). The strong internal magnetic field is the ultimate energy reservoir. It may twist the external magnetic field 
(Thompson et al. 2002). The balance between external untwisting and internal twisting may result in a relatively stable particle outflow. 
The nebula magnetic field may be related with the central magnetar's magnetic field (Shapiro \& Teukolsky 1983). However, in the case of magnetars, there are some differences with the normal pulsar case. Including the corresponding corrections, it is 
\begin{equation}
B_{\rm nebula} = B_{\rm ns}\times \left( \frac{R_{\rm ns}}{R_{\rm open}} \right)^{2+n} \times \left( \frac{R_{\rm open}}
{R_{\rm nebula}}\right),
\end{equation}
where $B_{\rm nebula}$ and $R_{\rm nebula}$ are the nebula magnetic field and radius, respectively; $B_{\rm ns}$ and $R_{\rm ns}$ are the magnetar's surface magnetic field and radius, respectively; $R_{\rm open}$ is the magnetic field line opening radius;  $n$ is a parameter describing the deviation of magnetic field from the dipole case. The surface magnetic field of magnetars depends on the braking mechanism etc. The magnetic field line opening radius may be equal to the light cylinder radius. The presence of particle wind will result in a smaller opening radius (Harding et al. 1999; Tong et al. 2013). The magnetic field in the magnetosphere of magnetars may be twisted compared with the dipole case (Wolfson 1995; Thompson et al. 2002). The parameter $n$ is in the range of $0 <n \le1$. 
The net effect is a higher magnetic field at the opening radius. With all these effects considered, a nebula magnetic field of $\sim 10^{-4} \,\rm G$ can be obtained. The detailed calculations of particle wind outflow and nebula magnetic field, the structure and evolution of a magnetar wind nebula will be the topic of further studies. 

In conclusion, it is shown that both the magnetar Swift J1834.9$-$0846 and its possible wind nebula are understandable. Using three different approaches, a particle luminosity of $10^{36} \,\rm erg \,s^{-1}$ to $10^{38} \,\rm erg \,s^{-1}$ is obtained. 
The nebula magnetic field is about $10^{-4} \,\rm G$. 
The possible wind nebula of Swift J1834.9$-$0846 should be a magnetar wind nebula. 

\section*{Acknowledgments}
The author would like to thank Y. Chen, P. Zhou for discussions. 
H.Tong is supported West Light Foundation of CAS (LHXZ201201), 973 Program (2015CB857100) 
and Qing Cu Hui of CAS.

\end{document}